\documentclass[conf]{IEEEtran}
\usepackage{graphicx}
\usepackage{fancyhdr}
\usepackage{makeidx}
\usepackage{amssymb,amsmath}
\usepackage{enumerate}
\newcommand{\bd}{\begin{document}}
\newcommand{\ed}{\end{document}}
\newcommand{\bc}{\begin{center}}
\newcommand{\ec}{\end{center}}
\newcommand{\vs}{\vspace}
\newcommand{\hs}{\hspace}
\newcommand{\beq}{\begin{equation}}
\newcommand{\eeq}{\end{equation}}
\newcommand{\beqs}{\begin{eqn*}}
\newcommand{\eeqs}{\end{eqn*}}
\newcommand{\bq}{\begin{quote}}
\newcommand{\eq}{\end{quote}}
\newcommand{\lb}{\linebreak}
\newcommand{\mb}{\makebox}
\newcommand{\fb}{\framebox}
\newcommand{\mc}{\multicolumn}
\newcommand{\ben}{\begin{enumerate}}
\newcommand{\een}{\end{enumerate}}
\newcommand{\bit}{\begin{itemize}}
\newcommand{\eit}{\end{itemize}}
\newcommand{\ov}{\overline}
\newcommand{\un}{\underline}
\newcommand{\lt}{\left}
\newcommand{\rt}{\right}
\newcommand{\ba}{\begin{array}}
\newcommand{\ea}{\end{array}}
\newcommand{\beqa}{\begin{eqnarray}}
\newcommand{\eeqa}{\end{eqnarray}}
\newcommand{\beqas}{\begin{eqnarray*}}
\newcommand{\eeqas}{\end{eqnarray*}}
\newcommand{\bfg}{\begin{figure}}
\newcommand{\efg}{\end{figure}}
\newcommand{\pad}{\partial}
\newcommand{\nn}{\nonumber}
\newcommand{\la}{\leftarrow}
\newcommand{\ra}{\rightarrow}
\newcommand{\lgla}{\longleftarrow}
\newcommand{\lgra}{\longrightarrow}
\newcommand{\La}{\Leftarrow}
\newcommand{\Ra}{\Rightarrow}
\newcommand{\Lra}{\Leftrightarrow}
\newcommand{\Lgla}{\Longleftarrow}
\newcommand{\Lgra}{\Longrightarrow}
\renewcommand{\a}{\alpha}
\renewcommand{\b}{\beta}
\newcommand{\g}{\gamma}
\newcommand{\G}{\Gamma}
\renewcommand{\d}{\delta}
\newcommand{\D}{\Delta}
\newcommand{\e}{\epsilon}
\newcommand{\eps}{\epsilon}
\newcommand{\s}{\sigma}
\renewcommand{\l}{\lamda}
\newcommand{\m}{\mu}
\newcommand{\n}{\nu}
\renewcommand{\S}{\Sigma}
\newcommand{\p}{\pi}
\newcommand{\om}{\omega}
\newcommand{\Om}{\Omega}
\newcommand{\tri}{\triangle}
\newcommand{\ti}{\times}
\newcommand{\f}{\frac}
\newcommand{\ds}{\displaystyle}
\newcommand{\bm}[1]{\mb{{\boldmath $#1$}}}
\newcommand{\alter}[2]{\lt\{ \ba{ll}#1 \\ #2 \ea \rt.}
\newcommand{\alt}[4]{\lt\{ \ba{ll}#1 & \mb{if \, \,}#2 \\ #3 & \mb{}#4 \ea
    \rt.}
\newcommand{\altn}[4]{\lt\{ \ba{rl}#1 & \mb{if \, \,}#2 \\ #3 & \mb{}#4 \ea
    \rt.}
\newcommand{\altif}[4]{\lt\{ \ba{ll}#1 & \mb{if \, \,}#2 \\ #3 &
\mb{if \, \,}#4 \ea \rt.}
\newcommand{\altnif}[4]{\lt\{ \ba{rl}#1 & \mb{if \, \,}#2 \\ #3 &
\mb{if \, \,}#4 \ea \rt.}
\newcounter{algc}
\newcounter{romc}
\newcounter{Alphc}
\newcommand{\bl}{\begin{list}{{\it Step} ~\arabic{algc}~:} {\usecounter{algc}
                \setlength{\topsep}{0pt} \setlength{\itemsep}{0pt}}}
\newcommand{\el}{\end{list}}
\newcommand{\blr}{\begin{list}{~\roman{romc}~:} {\usecounter{romc}
                \setlength{\topsep}{0pt} \setlength{\itemsep}{0pt}}}
\newcommand{\elr}{\end{list}}
\newcommand{\bla}{\begin{list}{~\Alph{Alphc}~:} {\usecounter{Alphc}
                \setlength{\topsep}{0pt} \setlength{\itemsep}{0pt}}}
\newcommand{\ela}{\end{list}}

\newtheorem{theorem}{Theorem}

\begin{document}
\title{\vspace{-0.3in}\Large{\textbf{High On-Off Ratio Bilayer Graphene Complementary Field Effect Transistors}}}
\author{Kausik Majumdar$^1$, Kota V. R. M. Murali$^2$, Navakanta Bhat$^1$, Fengnian Xia$^3$ and Yu-Ming Lin$^3$\\
\small{$^1$Department of Electrical Communication Engineering and
Center of Excellence in Nanoelectronics, \\Indian Institute of
Science, Bangalore 560012, India.\\ $^2$IBM Semiconductor Research
and Development Center, Bangalore
560045, India.\\
$^3$IBM T. J. Watson Research Center, Yorktown Heights, NY 10598,
USA.\\
\vspace{0.1in} Tel: +91-988-680-3566, Fax: +91-80-2360-0563, Email:
kausik@ece.iisc.ernet.in }}
\date{}
\maketitle

\bc
\textbf{Abstract}
\ec

In this paper, we propose a novel S/D engineering for dual-gated
Bilayer Graphene (BLG) Field Effect Transistor (FET) using doped
semiconductors (with a bandgap) as source and drain to obtain
unipolar complementary transistors. To simulate the device, a
self-consistent Non-Equilibrium Green's Function (NEGF) solver has
been developed and validated against published experimental data.
Using the simulator, we predict an on-off ratio in excess of $10^4$
and a subthreshold slope of $\sim$$110$mV/decade with excellent
scalability and current saturation, for a $20$nm gate length
unipolar BLG FET. However, the performance of the proposed device is
found to be strongly dependent on the S/D series resistance effect.
The obtained results show significant improvements over existing
reports, marking an important step towards bilayer graphene
logic devices.\\

\bc
\textbf{Introduction}
\ec

After the first isolation of Graphene on an oxide substrate, it has
attracted a tremendous amount of interest among the researchers due
to its extraordinary properties \cite{nov04}. Using Graphene as the
channel material of a FET, that eventually outperforms Si MOSFET,
has been one of the major interests in the recent years
\cite{lemme07}-\cite{sz10}. However, in spite of excellent
electronic properties \cite{nov04}, suitability for planar MOSFET
technology \cite{lemme07,lin10} and perfect 2-D electrostatics
\cite{sz10}, large area graphene is yet to compete with the existing
Si technology in CMOS VLSI due to a lack of significant bandgap,
which leads to leaky devices. With the discovery of moderate bandgap
opening by the application of a vertical field in a dual-gated BLG
structure \cite{to06,yz09}, it has become possible to improve the
off state leakage, though the on-off ratio of such transistors are
still significantly inferior to Si devices \cite{fx10}-\cite{gf09}.
In addition to the poor on-off ratio, the small bandgap in the
Graphene channel causes a number of issues that need to be overcome
before being used as logic devices, namely: (1) There is a lack of
sufficient drain current saturation \cite{sz10,fx10,im08}; (2) To
control the excessive leakage through the drain barrier, it is
required to operate the transistor at low drain bias condition,
which in turn adversely affects the on current \cite{fx10}; and (3)
The unipolar devices in Graphene (P-FET and N-FET), which are
required to obtain complementary operation in logic circuits, are
generally obtained by using charge neutrality point splitting and
hence are not very \emph{well-behaved} \cite{ft09,sl10}. In order to
address these issues, in the following, we propose a doped
semiconductor S/D engineering for BLG FETs which significantly
reduces either eletron or hole injection from the S/D contacts by
the choice of doping allowing us to obtain unipolar
high on-off ratio devices.\\

\bc
\textbf{Proposed Design}
\ec

\begin{figure}[h]
\hspace{0.1in}
\vs{-0.0in}
\includegraphics[scale=0.4]{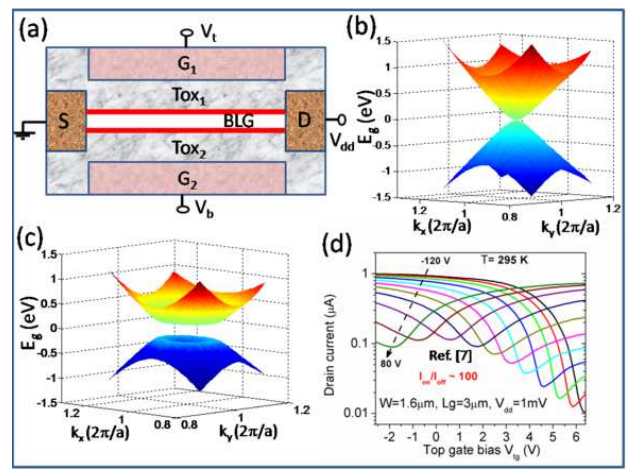}
\vs{0.1in} \caption{(a): The schematic of a dual-gated (G$_1$,
G$_2$) bilayer graphene FET with source (S) and drain (D). (b):
Bandstructure of an unbiased infinite BLG film with zero bandgap.
(c): Bandgap opening in a \emph{`Mexican hat'} shape under applied
external vertical field. (d): Typical experimental transfer
characteristics of a metal S/D BLG FET reported in ref. \cite{fx10}
with on-off ratio of $\sim$$100$ at $V_{dd}$=$1$mV and
$T$=$295$K.}\label{fig:schematic}
\end{figure}

The schematic diagram of a conventional dual-gated BLG FET is shown
in Fig. \ref{fig:schematic}(a) where the two gates (G$_1$, G$_2$)
can be used to control the bandgap and charge in the BLG channel
connecting the metal source and drain. As shown in Fig.
\ref{fig:schematic}(b) and (c), the BLG channel does not posses any
bandgap intrinsically, but a bandgap can be introduced by applying a
field through the top and the back gates. A typical transfer
characteristics obtained in such a transistor is shown in Fig.
\ref{fig:schematic}(d) \cite{fx10}.

\begin{figure}[h]
\hs{0.35in}
\includegraphics[scale=0.6]{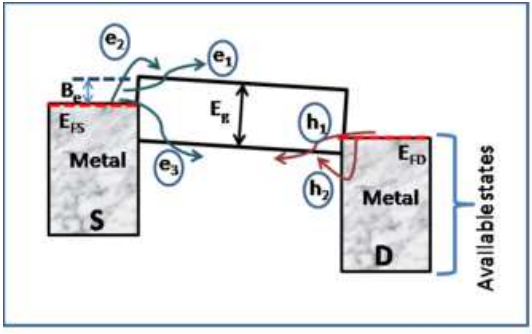}
\vs{-2in} \caption{Schematic band diagram of a conventional metal
S/D BLG FET during off state with major leakage paths indicated by
arrows. $e_1$: direct S/D tunneling of electrons; $e_2$: thermal
emission of electrons over source barrier; $e_3$: source junction
tunneling leakage of electrons; $h_1$: drain junction tunneling
leakage of holes; $h_2$: thermal emission of holes over drain
barrier through the available states in the metal. Simultaneous
reduction of electron and hole injection to channel is difficult in
this device which results in poor on-off current
ratio.}\label{fig:band_diagram1}
\end{figure}
The major off state leakage current paths in such a conventional
metal S/D BLG FET are shown in Fig. \ref{fig:band_diagram1}. Since
the bandgap ($E_g$) opened in the channel during off state is not
large enough, the source barrier for an electron, which is nearly
$E_g/2$, is also small. The barrier for the holes at the drain side
is thus even less due to presence of the drain bias $V_{dd}$. Hence
it is difficult to simultaneously reduce the leakage current through
both the source barrier and the drain barrier, resulting in poor off
state leakage.

\begin{figure}[t]
\hs{0.35in}
\includegraphics[scale=0.6]{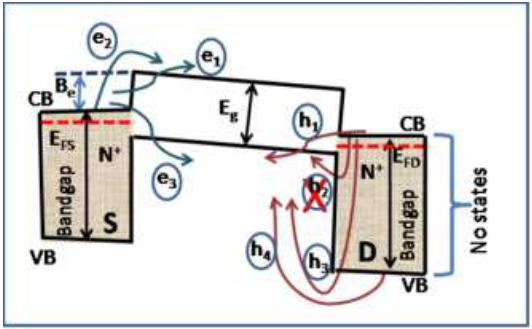}
\vs{-2in} \caption{The leakage paths in the proposed (N$^+$) doped
S/D BLG FET during off state: The thermal emission, $h_2$, which is
generally the major hole leakage component from the drain contact,
is no more allowed due to lack of states in the bandgap of the
drain. However, the thermal emission $h_3$ over the bandgap is
possible, but is negligible due to the large energy required by the
process. Leakage from drain valence band ($h_4$) is also negligible
due to the lack of holes in the valence band of the N$^+$ doped
drain. Reduced hole leakage from the drain in turn allows us to
optimize the biasing condition to increase the source barrier $B_e$
reducing $e_1$ and $e_2$. However, increasing $B_e$ too much
increases tunneling leakage $e_3$ and $h_1$ adversely affecting
overall off state leakage. We can similarly obtain a BLG P-FET using
P$^+$ doped S/D. }\label{fig:band_diagram2}
\end{figure}

In the proposed solution, as explained in Fig.
\ref{fig:band_diagram2}, the metal source and drain are replaced by
an N$^+$ (or P$^+$) doped semiconductor with a bandgap to obtain a
BLG N-FET (or P-FET). We observe in Fig. \ref{fig:band_diagram2}
that the lack of states in the bandgap of the N$^+$ doped drain
significantly reduces the major hole leakage paths. In particular,
the thermal hole injection (path $h_2$) through the drain barrier is
completely switched off as the unavailability of states in the
bandgap of the drain does not allow this process. The leakages
through paths $h_3$ and $h_4$ are also almost negligible. The
significant reduction of the hole injection from the drain in turn
allows us to choose the gate biasing condition such that the
electron barrier at the source ($B_e$) can be significantly
increased, helping to suppress the leakage through paths $e_1$ and
$e_2$ as well.
\begin{figure}[hbt!]
\vs{-0.05in}
\hs{0.1in}
\includegraphics[scale=0.4]{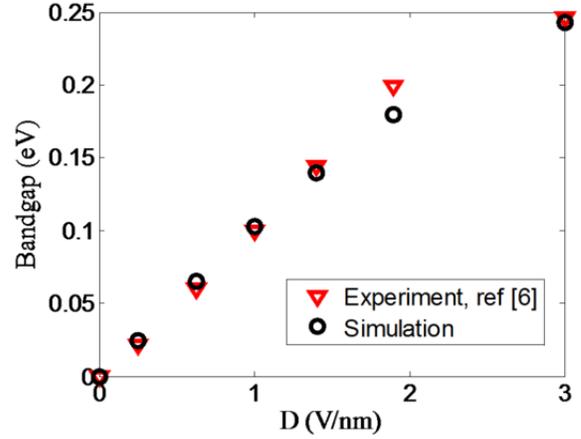}
\vs{-0.1in} \caption{The simulated external field dependent bandgap
of a dual gated BLG film as a function of the displacement vector
defined in a similar way as in \cite{yz09}. There is an excellent
agreement with the experimental results reported in
\cite{yz09}.}\label{fig:bandgap_match}
\end{figure}
However, excessive pulling up of the source barrier will increase
the junction tunneling leakages through paths $e_3$ and $h_1$
adversely affecting the total leakage current. Note that, in case of
conventional metal S/D, this source barrier pulling will aggravate
the hole injection from the drain contact and hence can not be used.
The reduced hole injection from the drain in the proposed scheme
also allows us to operate the BLG FET at higher drain bias,
significantly improving the drive current at a given off state
current, as compared with metal S/D BLG FETs \cite{fx10}.
Interestingly, the proposed P-FET and N-FET designs are almost
symmetric with each other due to the inherent symmetry of the
conduction band and the valence band of the BLG channel - an
advantage over Si technology. However, in reality, non-identical
barrier heights between the BLG channel and the P-type or N-type
semiconductor in the source and the drain can introduce asymmetry
between the P-FET and the N-FET.\\

\bc
\textbf{Simulation Method and Validation}
\ec

\begin{figure}[hbt!]
\hs{0.1in}
\includegraphics[scale=0.4]{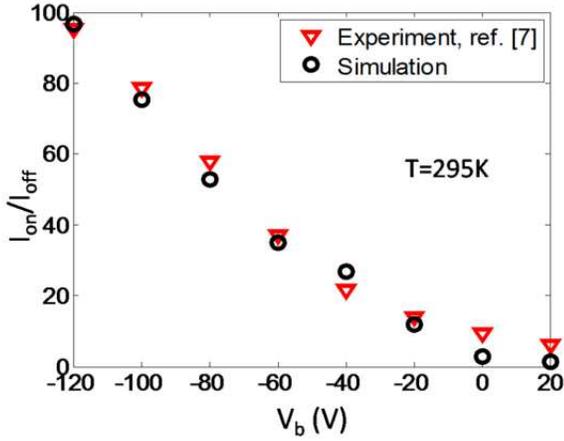}
\caption{Simulated on-off current ratio as a function of
the back gate bias ($V_b$) in a similar device as in \cite{fx10}
with $L_g$=$1.6\mu$m and Ti/Pd/Au/Ti S/D at $1$mV drain bias. The on
state is defined at $V_t$=$-2.6~$V and off states are the charge
neutrality points. The ratios are in good agreement with the
experimental results. The minor mismatches can be attributed to the
ballistic assumption of the channel in the simulation ignoring
scattering. This calibration of the simulation method forms the
basis of the results presented in the rest of the
paper.}\label{fig:onoff_match}
\end{figure}
\begin{figure}[hbt!]
\hs{0.1in}
\includegraphics[scale=0.4]{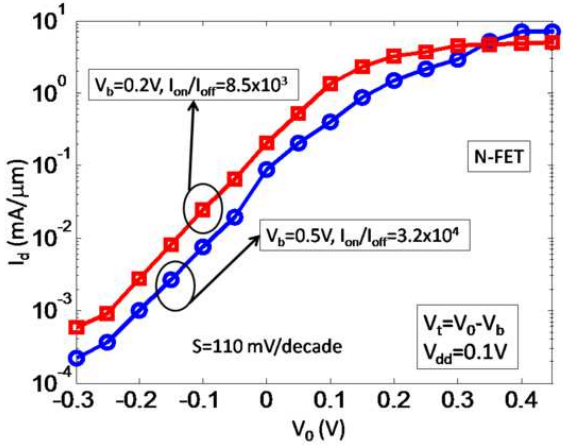}
\caption{Transfer characteristics of the proposed BLG
FET (N-type) with $20$nm gate length, $1$nm top and back gate EOT
and $0.1$V drain bias keeping the bottom gate bias at $0.5$V and
$0.2$V. The former case provides better on-off ratio due to
increased bandgap. Both the on-off ratio ($3.2\times10^4$) and the
subthreshold slope (S) ($110~$mV/decade) are the best reported to
date for a BLG FET.}\label{fig:IdVg_log}
\end{figure}
\begin{figure}
\hs{0.1in}
\includegraphics[scale=0.4]{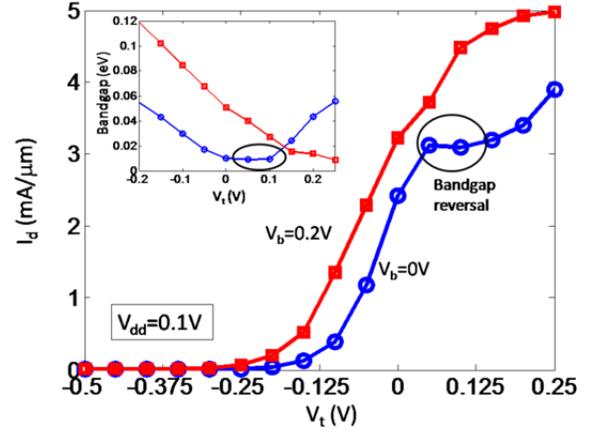}
\caption{Transfer characteristics of the proposed BLG FET (N-type)
in the linear scale with the back gate kept at $0$V and $0.2$V. In
the former case, a valley is observed in the characteristics due to
bandgap reversal effect. Inset: The change in bandgap of the BLG
channel as a function of the top gate bias for the two
cases.}\label{fig:IdVg_linear}
\end{figure}
We have developed a self-consistent NEGF \cite{yo06,gf09} solver to
simulate the proposed dual-gated BLG FET. The device Hamiltonian is
constructed by discretizing the real space along the channel length,
but a Bloch periodicity is assumed along the width of the channel.
The effects of the source and the drain are captured using
appropriate self-energy matrices \cite{sd}. At every iteration, once
the charge distribution is obtained using the LDOS and Fermi
statistics, the self-consistent potential is found by solving
Poisson equation with appropriate boundary conditions \cite{km10}.
The real space NEGF simulator takes care of both the thermal as well
as the tunneling components of the total drain current. To validate
the simulator, in Fig. \ref{fig:bandgap_match}, we find close
agreement between the simulated external field dependent bandgap in
a BLG film and the corresponding experimental data reported in
\cite{yz09}.

We also obtained the on-off current ratios in a metal S/D BLG FET
with similar device parameters as in \cite{fx10}. The simulation
predicted numbers are found to be in reasonable agreement with the
experimental data, as shown in Fig. \ref{fig:onoff_match}. In the
rest of the paper, we assume a $20$nm gate length BLG FET with $1$nm
equivalent oxide thickness (EOT) at the top and back gates having
zero offset bias. The doped source and drain are assumed to have
a bandgap of $1.1$eV and perfect interface with the channel.\\

\bc \textbf{Results and Discussions} \ec

\begin{figure}
\hs{0.1in}
\includegraphics[scale=0.4]{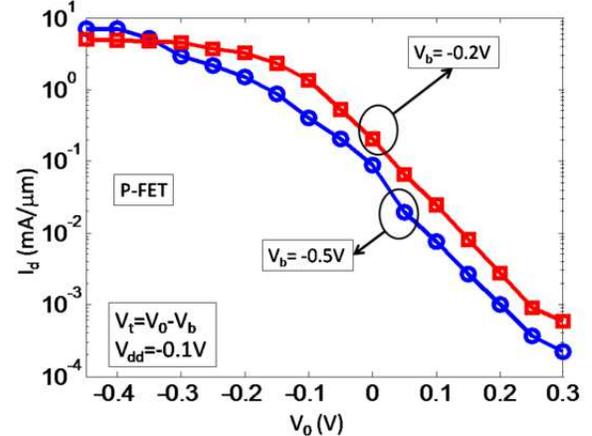}
\vs{-0.2in} \caption{Transfer characteristics of the proposed BLG
FET (P-type) with $20$nm gate length and $-0.1$V drain bias keeping
the back gate at $-0.5$V and $-0.2$V. The characteristics is similar
to the N-FET arising from inherent symmetry of the conduction band
and the valence of the BLG channel. This is an advantage over the
existing Si technology where symmetry between NMOS and PMOS is
obtained by sizing of transistors.}\label{fig:IdVg_log_p}
\end{figure}

The transfer characteristics of the proposed BLG N-FET are plotted
in Fig. \ref{fig:IdVg_log} which show an on-off ratio in excess of
$10^4$ with a subthreshold slope of nearly $110~$mV/decade. Both of
these are significant improvements over existing experimental data
\cite{fx10} as well as theoretical predictions \cite{km10} for metal
S/D BLG FET. This is primarily due to the significant reduction of
hole injection through the drain barrier. The transfer
characteristics are plotted in the linear scale in Fig.
\ref{fig:IdVg_linear} which clearly shows a valley for $V_b$=$0$.
This is essentially due to the fact that the rate of change of the
BLG bandgap changes its sign close to $V_t$=$0$, as shown in the
inset. In Fig. \ref{fig:IdVg_log_p}, we plot the transfer
characteristics for a BLG P-FET. The observed similarity between the
P-FET and the N-FET characteristics reflects the quasi-symmetric
conduction band and valence band in the Graphene channel. Here it is
important to note that we have assumed similar barrier height
characteristics between the BLG channel and the P-type or N-type
doped semiconductor S/D.

\begin{figure}
\hs{0.1in}
\includegraphics[scale=0.4]{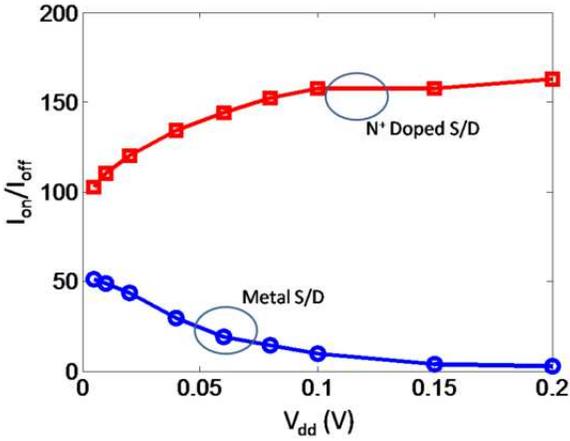}
\vs{-0.1in} \caption{On-Off ratio as a function of drain bias shows
significant degradation of metal S/D BLG FET due to an increase in
the hole injection from the drain contact. In case of the doped
N$^+$ S/D, the problem is resolved by the bandgap of the drain and
we observe an increase in the on-off ratio with drain bias. Ability
to operate the FET at larger drain bias in turn improves the on
current significantly. In both cases, the on and off conditions are
defined as $V_t=V_b=0.5$V and $V_t=-V_b=0.5$V respectively. Note
that, the off state is defined in this way for a fair comparison
between the two cases and is certainly not optimum for the doped S/D
FET.}\label{fig:onoff_Vd}
\end{figure}

\begin{figure}[h]
\hs{0.1in}
\includegraphics[scale=0.4]{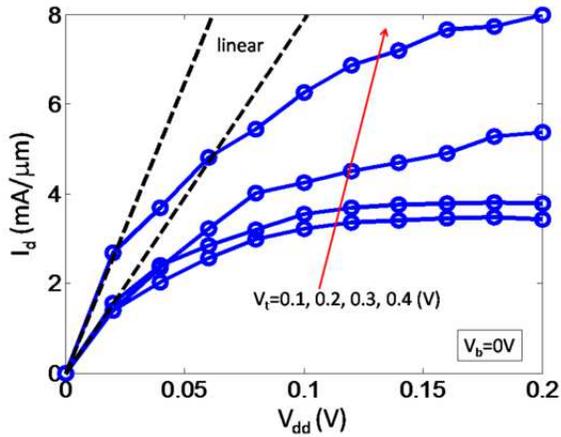}
\vs{-0.1in} \caption{Output characteristics of the proposed BLG FET
show improved saturation of drain current with drain bias. The
dotted lines represent the resistor characteristics (linear Ohm's
law).}\label{fig:IdVd}
\end{figure}

In Fig. \ref{fig:onoff_Vd}, we notice that with an increase in drain
bias, there is a drastic degradation in the on-off ratio in the case
of metal S/D due to significant lowering of the drain barrier height
for holes (path $h_2$ in Fig. \ref{fig:band_diagram1}). This forces
us to operate the FET at low drain bias which in turn degrades the
drive current. However, this issue can be potentially solved by
using doped S/D which eliminates leakage through path $h_2$ anyway
(Fig. \ref{fig:band_diagram2}). The output characteristics in Fig.
\ref{fig:IdVd} show good saturation of the drain current and this
consequently enable us to improve gain and noise margin in digital
logic as well as gain in RF applications.

\begin{figure}[h]
\includegraphics[scale=0.43]{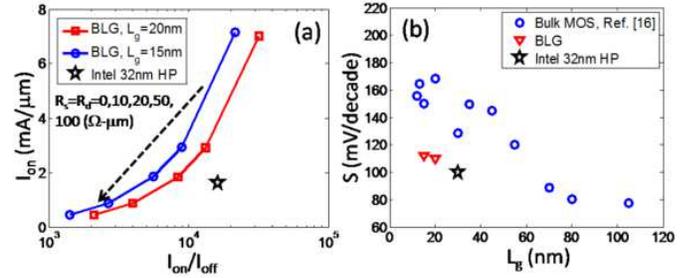}
\vs{-1.4in} \caption{(a): The on current of the proposed device
plotted as a function of on-off ratio shows excellent scalability
down to $15$nm gate length and compares very well with state of the
art Si MOSFET technology. The parasitic series resistances at S/D
($R_{s,d}$) are added for realistic performance comparison.
Reduction of $R_{s,d}$ plays a key role in maintaining the high
drive current provided from the intrinsic device. (b): Subthreshold
slope ($S$) remains almost the same for both $20$nm and $15$nm gate
length device showing scaling potential of the proposed device due
to its ultra-thin structure. However, driving $S$ closer to
Boltzmann limit is difficult due to the intrinsic bias dependence of
the BLG bandstructure \cite{km10}. This scalability is promising
when compared with existing state of the art Si technology
\cite{rc05}.}\label{fig:perfm}
\end{figure}

In Fig. \ref{fig:perfm}(a), the on current is plotted against on-off
current ratio for a gate length of $15$nm and $20$nm of the proposed
device and is compared against state of the art Si technology. The
S/D series resistances are added for realistic performance
evaluation and are found to play a key role in determining both the
drive current and the on-off ratio.
Note that, unlike Si MOSFET, the limit of the subthreshold slope
($S$) in the proposed BLG FET primarily arises from the intrinsic
bias dependence of the electronic structure of BLG \cite{km10}, and
not much from the gate length scaling, due to its ultra-thin
structure. This fact plays an important role in determining the
scalability of the proposed BLG transistor. This is reflected in
Fig. \ref{fig:perfm}(b), where the plotted subthreshold slopes,
though are well above the Boltzmann limit, show negligible
degradation with gate length providing excellent scalability. The
computed subthreshold slopes are found to compare very well with Si
MOSFET technology.\\

\bc
\textbf{Conclusion}
\ec

In conclusion, we have proposed a doped semiconductor S/D
engineering scheme to obtain complementary unipolar BLG FETs. A
self-consistent NEGF simulator has been developed and validated
against published experimental data to predict the device
characteristics. The obtained unipolar characteristics show
significant improvement of on-off current ratio, subthreshold slope
and drain current saturation over existing published data. The
proposed transistor characteristics compare well with state of the
art Si technology $-$ marking an important step toward bilayer
graphene
logic devices.\\

\bc \textbf{Acknowledgement} \ec

K. Majumdar and N. Bhat would like to thank the Ministry of
Communication and Information Technology, Government of India, and
the Department of Science and Technology, Government of India, for
their support.

\end{document}